\documentclass[reprint,prl,aps,superscriptaddress,showkeys,showpacs]{revtex4-1}
\usepackage{graphicx,times,CJK}
\begin{document}
\begin{CJK*}{UTF8}{}
\title{Drying of complex suspensions}

\author{Lei Xu (\CJKfamily{bsmi}徐磊)}
\affiliation{Department of Physics, The Chinese University of Hong Kong, Hong Kong}
\affiliation{Department of Physics and SEAS,
Harvard University, Cambridge, MA 02138, USA}
\author{Alexis Berg\`es}
\affiliation{D\'epartement de physique, \'Ecole Normale
Sup\'erieure, Paris, France}
\author{Peter J.~Lu (\CJKfamily{bsmi}陸述義)}
\affiliation{Department of Physics and SEAS, Harvard
University, Cambridge, MA 02138, USA}
\author{Andr\'e R.~Studart}
\affiliation{Complex Materials, Department of Materials, ETH Zurich, 8093 Zurich, Switzerland}
\author{Andrew B. Schofield}
\affiliation{The School of Physics, University of Edinburgh,
Edinburgh EH9 3JZ, UK}
\author{Hidekazu Oki}
\affiliation{Shinagawa-ku, Tokyo, Japan}
\author{Simon Davies} \affiliation{AkzoNobel Corporate,
R320 Wilton Centre, Redcar TS104RF, UK}
\author{David A.~Weitz}
\affiliation{Department of Physics and SEAS, Harvard
University, Cambridge, MA 02138, USA}

\begin{abstract}
We investigate the 3D structure and drying dynamics of complex
mixtures of emulsion droplets and colloidal particles, using confocal microscopy.
Air invades and rapidly collapses large emulsion droplets, forcing their contents
into the surrounding porous particle pack at a rate proportional to the square of
the droplet radius. By contrast, small droplets do not collapse, but
remain intact and are merely deformed. A simple model coupling the Laplace
pressure to Darcy's law correctly estimates both the threshold radius separating
these two behaviors, and the rate of large-droplet evacuation. Finally, we use these
systems to make novel hierarchical structures.
\end{abstract}
\maketitle
\end{CJK*}
The drying of suspensions of colloidal particles gives rise a
plethora of fascinating phenomena, from the ``coffee-ring''
effect\cite{Deegan} to episodic crack
propogation\cite{Dufresne} and the fractal patterns arising
from invasion percolation\cite{Xu,Wilkinson,Page,Shaw,Robbins1}.
Drying of colloidal suspensions is also
important technologically: paints and other coatings depend on
colloidal particles for many of their key properties, many
ceramics go through a stage of particle drying, and cosmetics
often exploit the unique properties of colloidal-scale
particles, particularly for such beneficial properties as
screening the harmful effects of the sun. However, for many of
these technological applications, the colloidal particles are
but one of many different components, and drying of the
colloids is accompanied by many other phase changes. While
these mixtures can become highly complex, a simpler, yet still
rich system that embodies many of the complex phenomena of
these technological suspensions is a mixture of immiscible
fluids with a colloidal suspension; a simple example is a
mixture of an emulsion and colloidal particles. The behavior of
the emulsion embodies many of the archetypal phenomena of such
systems, while still remaining sufficiently tractable to enable
it to be fully understood. However, emulsions themselves
typically scatter light significantly, and when mixed with a
colloidal suspension, this scattering is only enhanced. As a
result, it is very difficult to image this mixture, precluding
optical studies of its behavior, and knowledge of the actual
behavior is woefully missing.

In this Letter, we explore the drying of mixtures of aqueous emulsion droplets
and spherical colloidal particles with confocal microscopy, which allows us to
resolve the full 3D structure of these mixtures and their temporal
dynamics. We find that the particles first jam into a solidified pack, throughout
which emulsion drops are dispersed; a front of air then passes through the entire
system. When this drying front reaches large emulsion droplets, the droplets
unexpectedly collapse and their internal contents are forced into the pore space
between the surrounding colloids, driven by an imbalance of pressures at the
droplets' interfaces with air and with the solvent. By contrast, small droplets
are deformed by the drying front, yet remain intact without bursting. By coupling
the Laplace pressure with Darcy's law for flow through a porous medium, we
predict the duration of large droplet invasion, and show that the threshold size
between bursting and deformation is comparable to the size of the colloidal
particles. We use this technique to create novel hierarchical materials.
\begin{figure}
\begin{center}
\includegraphics[width=2.9in]{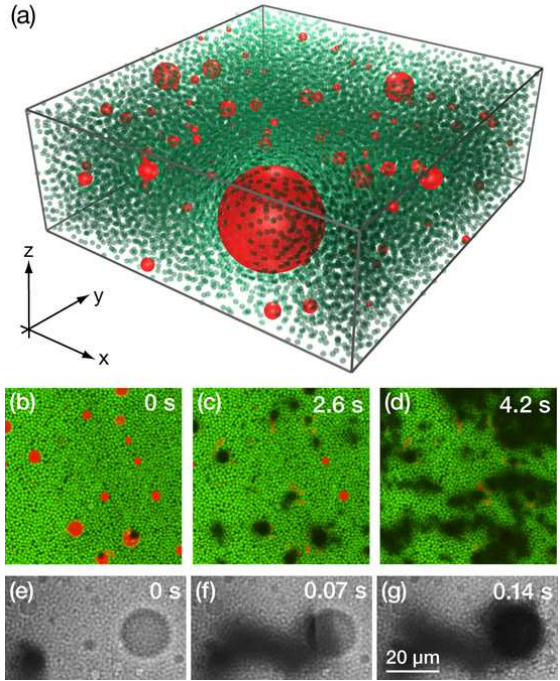}
\caption{
Structure and drying of droplet-particle mixtures. (a) Reconstruction of a
typical sample from confocal microscope images, with dimensions $91 \times 91
\times 30$ $\mu$m$^3$. Polydisperse red spheres are emulsion droplets;
monodisperse green spheres are PMMA particles, shown at half size for clarity.
(b)-(d) Two-dimensional confocal microscope images in $x$-$y$ plane, within the
bulk of the sample, showing the invasion of the drying front. (b) Onset of air
invasion. (c) As the drying front moves through, emulsion droplets turn black
first, followed by (d) air invading the particle regions. (e)-(g) Bright-field
images showing droplet invasion, where air appears black and the solvent is
transparent. Air (e) approaches, (f) contacts, and (g) rapidly evacuates the
droplet.}
\end{center}
\end{figure}

We suspend sterically-stabilized colloidal spheres of
polymethylmethacrylate (PMMA) with radius $r_\mathrm{p}=1$
$\mu$m in decahydronaphthalene (DHN). Separately, we create an
emulsion of an aqueous phase, comprising equal volumes of water
and glycerol, and PGPR-90 surfactant, required for droplet stability, in the
nonpolar DHN; our homogenizer creates polydisperse
droplets ranging from microns to tens of microns. We combine
particle suspension and emulsion to create our
particle-droplet mixtures. These particular components ensure
that the refractive indices of the particles, droplets and
background solvent are all sufficiently matched that we can
image the entire bulk of the 3D structure with confocal
fluorescence microscopy, with single-particle
resolution\cite{Peter}. To distinguish particles from
droplets, we use different dyes: particles are dyed with
nitrobenzoxadiazole (NBD) and appear green;
droplets are dyed with rhodamine-B and appear red.

We deposit the particle-droplet mixture on a clean glass coverslip, and image
with an inverted confocal microscope (Leica SP5). The droplets are coated
immediately by colloids, reminiscent of a pickering
emulsion\cite{Velev,Dinsmore}. As in the case of particle-only systems,
evaporation proceeds in two stages\cite{Xu}. First, particles and droplets are driven toward the edge of the
drying sample, where they jam, analogous to the coffee-ring effect in
particle-only systems\cite{Deegan}. Then, a drying air front invades the jammed
system and displaces the DHN, which ultimately evaporates completely; we do not
observe cracks during drying, as expected for particles of this size\cite{Xu}. We
use confocal microscopy to observe the particle-droplet mixture after jamming, as
the drying front passes through, which allows us to determine the exact size and
position of each emulsion droplet and colloidal particle\cite{Peter}. A 3D
reconstruction of a typical jammed mixture before drying is shown in Fig.~1(a),
with droplets and particles shown in red and green, respectively. Collecting this
full 3D data takes several seconds, far too slow to observe the rapid dynamics
that occur during air invasion; instead, to capture these dynamics, we fix the
focal plane in the bulk of the sample and collect high-speed 2D images every 70
ms. By imaging the same regions in 3D first, then rapidly in 2D as the air
invades, we can observe the drying process with good spatial and temporal
resolution.
\begin{figure}
\begin{center}
\includegraphics[width=2.9in]{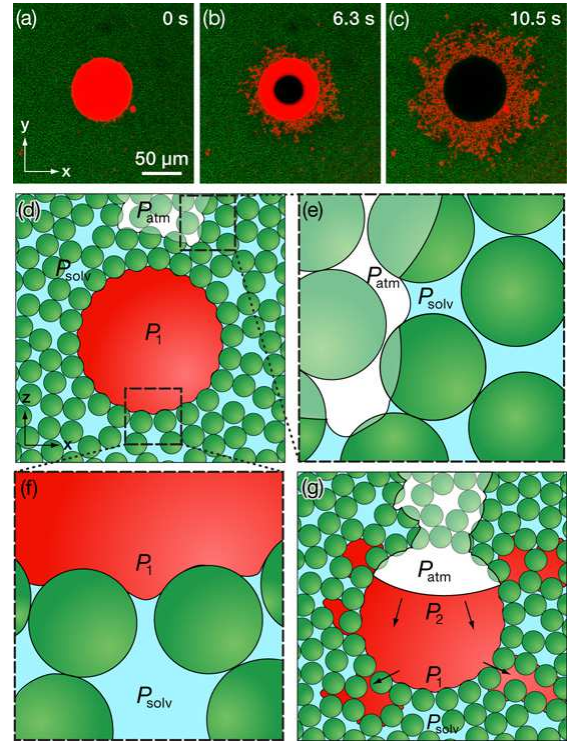}
\caption{
(a)-(c) Confocal microscope images of the air invasion of a large droplet viewed
at a fixed $x$-$y$ plane, where air forces the contents of the droplet into the
surrounding colloids. Before air invasion, tiny projections of the droplet extend
into the surrounding pores, visible in (a). (d) Schematic in the $x$-$z$ plane
showing the pressure distribution before air (white) touches the droplet.
Enlargement of the (e) air-solvent and (f) droplet-solvent interfaces. (g)
Schematic of the pressure distribution during evacuation. Inside the droplet, the
region next to air-droplet interface has a high pressure $P_2$; the region next
to solvent-droplet interface has a low pressure $P_1$. The difference between
these pressures drives the flow of droplet contents into the surrounding
particles.}
\end{center}
\end{figure}

In the jammed configurations before air invasion, the emulsion
droplets are distributed throughout the colloidal particles, as shown in
Fig.~1(a). As the sample dries, more and more particle regions turn black; the
invading air displaces the solvent around the particles, destroying the
refractive index match, as illustrated in Figs.~1(b)-(d). The droplets also turn
black as drying proceeds, but whether they are replaced by air or solvent cannot
be resolved with fluorescence, where both appear black. We therefore observe
droplet invasion with bright-field transmission microscopy, where the solvent
remains transparent but air appears black. We observe that droplets are replaced
by material that appears black in bright-field, which therefore must be air, as
shown in Figs.~1(e)-(g). For a more detailed understanding of droplet behavior,
we observe large isolated emulsion droplets, with radii $R_\mathrm{d}$ greater
than a few microns, during drying. Because it is energetically unfavorable to
displace the pre-existing organic solvent, which wets the PMMA particles, with
the aqueous droplet fluid that does not, we should expect that the aqueous fluid
remains in the droplet and slowly evaporates upon contact with air, with very
little fluid motion. Contrary to this expectation, however, we instead observe
significant fluid motion: air rapidly forces the contents of the droplet into the
pore space between the surrounding particles, leaving an empty, spherical void,
as shown in Figs.~2(a)-(c). This flow can not occur spontaneously; instead, a
strong driving force must exist.

To understand this driving force, we analyze the distribution of pressures inside
and around an isolated droplet. The total pressure near an interface is
determined by a combination of the external pressure and the interface's Laplace
pressure. The magnitude of the Laplace pressure is usually estimated as the ratio
of the interfacial tension to interface's curvature radius; its sign positive for
convex interfaces, negative for concave. Before air invasion, the droplet is
surrounded by particles and solvent, as shown in Fig.~2(d). The air protrudes
into the solvent, making a concave profile with a negative Laplace pressure, as
shown in Fig.~2(e)\cite{Dufresne}. We estimate $P_\mathrm{solv}$, the pressure of
the solvent, as the difference between atmospheric pressure $P_\mathrm{atm}$, and
the Laplace pressure of the air-solvent interface, as shown in Fig.~2(e):
$P_\mathrm{solv} \cong P_\mathrm{atm} - \sigma_\mathrm{air|solv}/a$, where
$\sigma_\mathrm{air|solv}$ is the interfacial tension of the air-solvent
interface, and $a$ is the typical size of the pores between colloidal particles.
The tiny pore size produces large negative Laplace pressures, and hence a low
$P_\mathrm{solv}$. As a result, the droplet is surrounded by a low-pressure
environment.

At the interface between droplet and solvent, the aqueous
droplet protrudes into the non-polar solvent around the
colloids, as shown in Fig.~2(f); the Laplace pressure here is
therefore positive. We estimate the pressure inside the droplet
as $P_1 \cong P_\mathrm{solv} + \sigma_\mathrm{drop|solv} /a$,
where $\sigma_\mathrm{drop|solv}$ is the interfacial tension of
the droplet-solvent interface.

As soon as air touches the droplet, however, the pressure distribution changes
dramatically. In particular, part of the droplet is now in contact with the
much-higher atmospheric air pressure, as illustrated in Fig.~2(g). Near the
air-droplet interface, the pressure inside the drop is estimated as $P_2 \cong
P_\mathrm{atm} - \sigma_\mathrm{air|drop} / R_\mathrm{d}$, where we have
estimated the radius of air-droplet interface by the droplet size $R_d$. The
pressure near the droplet-solvent interface remains $P_1$, as illustrated in
Fig.~2(g). If $P_2 > P_1$, then the pressure difference will force droplet fluid
into the surrounding pore space between colloidal particles. We estimate this
pressure difference:
\begin{equation}
\Delta P = P_2 - P_1 \cong \frac{\sigma_\mathrm{air|solv}}{a} -
\frac{\sigma_\mathrm{air|drop}}{R_\mathrm{d}}
- \frac{\sigma_\mathrm{drop|solv}}{a} \label{eqn:Delta_P}
\end{equation}
where any change in solvent pressure across the droplet is negligible.
Interestingly, the $\Delta P$ depends not on atmospheric pressure, but rather
on the competition between the Laplace pressures at the
various interfaces. We measure the corresponding surface tensions with the
pendant drop method: $\sigma_\mathrm{air|solv} = 26 \pm$ 2 mN/m,
$\sigma_\mathrm{air|drop} = 51\pm 3$ mN/m, and $\sigma_\mathrm{drop|solv} = 3.8
\pm 0.3$ mN/m. Because of the presence of surfactant at the interface, the
droplet-solvent surface tension $\sigma_\mathrm{drop|solv}$ is so much lower than
the other two that its contribution to the overall pressure difference is
negligible. The radii that we measure for the large droplets, $R_d \cong 10 - 50$
$\mu$m, is orders of magnitude larger than the size of the inter-particle pore
space, $a \cong 0.36 r_\mathrm{p} = 0.36$ $\mu$m for random close packed
particles\cite{Frost}. Consequently, the contribution from a pressure drop across
the air-droplet interface is also small, and the flow is essentially driven by
the low pressure in the solvent evacuating the large drops, as shown in
Figs.~2(a)-(c). Since the solvent strongly wets the particles, the menisci of the
air evaporating the solvent from the pores creates a low pressure in the
solvent. It is this, reflected in the first term in Eqn.~\ref{eqn:Delta_P}, which
establishes the large pressure difference that drives the flow. We estimate
$\Delta P \cong 0.6$ atm for these large drops.
\begin{figure}
\begin{center}
\includegraphics[width=3in]{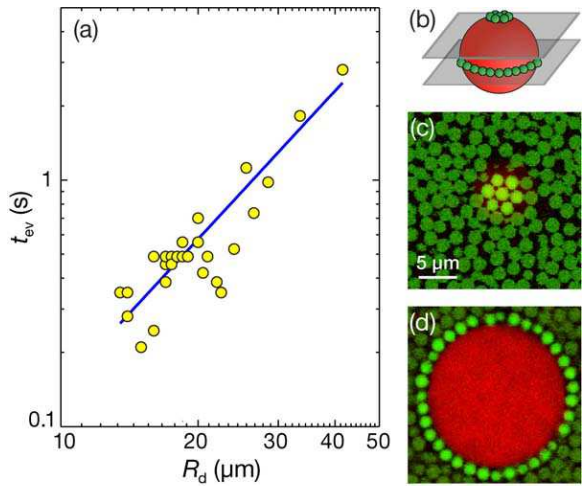}
\caption{ (a) Measurement of droplet evacuation duration time,
$t_\mathrm{ev}$, as a function of droplet radius, $R_\mathrm{d}$,
for a typical drying experiment. For the measured half decade, the data are consistent with a
power-law of slope 2, confirming the prediction of
eqn.~\ref{eqn:evac_time}. (b) Schematic of 2D images representing
slices through a droplet at different heights. (c) Confocal
microscope image at the top of the droplet, showing the hexagonal
crystalline ordering of the particles that suggests $\phi \approx  0.74$. (d)
Confocal microscope image through the droplet center, showing the dense packing around the droplet, consistent
with  high $\phi$. In these images, the brightness of
the particles in contact with the droplet has been enhanced for
clarity.}
\end{center}
\end{figure}

The high time resolution afforded by the rapid collection of 2D
image sequences in the confocal microscope allows us to measure
$t_\mathrm{ev}$, the evacuation time for air to invade the
large droplets and force their contents into the surrounding
particles; $t_\mathrm{ev}$ varies for droplets of different
sizes. We measure the size of each droplet from the initial 3D
confocal data, then quantify $t_\mathrm{ev}$ using fast 2D
images of droplet evacuation. Collected from deep in the bulk
of the sample, our images are large enough to contain a number
of droplets to obtain good statistics, yet are small enough
relative to the sample size to achieve a uniform sampling
environment.We find that the variation of $t_\mathrm{ev}$ with
$R_\mathrm{d}$ is consistent with a power-law, albeit over less than a decade. The
exponent is approximately 2, as shown by the comparison of the solid line
with the data points on the log-log plot in Fig.~3(a).

\begin{figure}
\begin{center}
\includegraphics[width=3.2in]{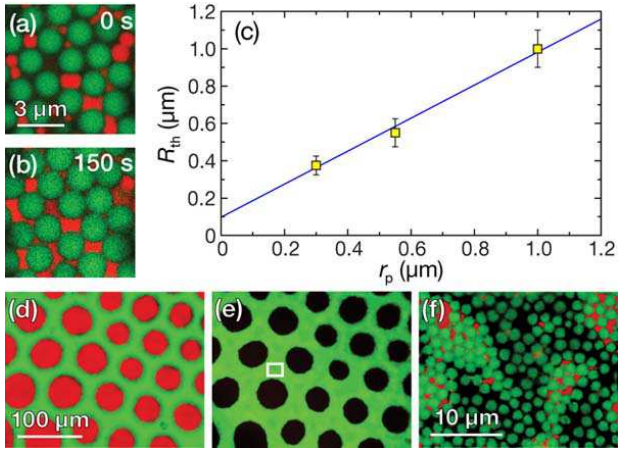}
\caption{
(a)-(c) Behavior of small droplets during drying. Small droplets
are (a) deformed but not invaded, and eventually (b) coalesce to
form a continuous phase. (c) The threshold droplet radius,
$R_\mathrm{th}$, vs. the particle radius, $r_\mathrm{p}$. A
linear fit through the data points,
$R_\mathrm{th}=(0.88\pm0.16)r_\mathrm{p}+(0.1\pm0.1)$,
consistent with our prediction
$R_\mathrm{th}=0.83r_\mathrm{p}$. (d)-(f) Hierarchical porous material created
from drying. Structure made from $R_d \cong 20$ $\mu$m droplets and $r_p = 1$
$\mu$m particles (d) before air invasion, and (e) after invasion. (f)
Enlargement of the compact region marked by the box in (e). The
particles appear non-spherical because of the difference in
their refractive index relative to air. The system contains
pores of two characteristic sizes: voids from droplets (20
$\mu$m) and inter-particle pores (0.5 $\mu$m).
}
\end{center}
\end{figure}

We estimate this relationship theoretically with a simple model
based on Darcy's Law for fluid flow in a porous
medium\cite{Batchelor} to determine a characteristic evacuation
velocity, $v_\mathrm{ev} = \kappa \nabla P / \mu$, where
$\kappa$ is the permeability of the porous medium, and $\mu$ is
the dynamic viscosity of the fluid. We estimate $\nabla P$, the
pressure gradient that drives the aqueous fluid into the porous
medium, as the characteristic pressure difference divided by
the droplet diameter, $\nabla P = \Delta P / 2 R_\mathrm{d}$.
We then estimate $t_\mathrm{ev} \cong 2R_\mathrm{d} / v_\mathrm{ev}$, yielding:
\begin{equation}
t_\mathrm{ev} = \frac{4\mu}{\kappa \Delta P} R_\mathrm{d}^2
\label{eqn:evac_time}
\end{equation}

The model predicts that $t_\mathrm{ev}$ data follow an $R_\mathrm{d}^2$
dependence; indeed, the experimental data closely conform to this particular
power-law scaling, as shown on the log-log plot in Fig.~3(a). Moreover, we can
further test the model by estimating the prefactor $4 \mu / (\kappa \Delta P)$ in
Eqn.~\ref{eqn:evac_time}. We estimate $\kappa$ using Kozeny-Carman equation for
flow through a porous medium\cite{Carman}, $\kappa \cong r_\mathrm{p}^2(1-\phi)^3
/ (45\phi^2)$, where the pore size is determined by the particle volume
fraction $\phi$. From the 3D particle positions in the bulk, we measure $\phi =
0.63 \pm 0.03$, consistent with random close-packing; however, around the large
droplets, the particles are hexagonally
close-packed\cite{Dinsmore}, and the next layer tightly packs the interstices,
as illustrated in Figs.~3(b)-(d). We think these well-ordered particles
predominantly determine the pore size through which the droplets evacuate. We
therefore estimate the permeability using the higher $\phi=0.74$. We measure $\mu = 8.9 \pm 0.1$ mPa-s for the H$_2$O/glyercol mixture, yielding our rough estimate of $4 \mu / (\kappa \Delta P) \approx 0.8 \times 10^9$ s/m$^2$; this value is of the same order as the
value of $t_\mathrm{ev} /R_\mathrm{d}^2= 1.2 \times 10^9$ s/m$^2$ from the fit to the
experimental data in Fig.~3(a), and provides support that our model
correctly captures the proper physics.

Our model explains the air invasion of large droplets, whose contents are forced
into the surrounding pore space. It also implies a completely different behavior
for sufficiently small droplets: as $R_\mathrm{d}$ decreases, $\Delta P$ must
also decrease, as can be seen from Eqn.~(\ref{eqn:Delta_P}). Eventually, for
sufficiently small drop size, $\Delta P$ will become zero and will no longer
drive any flow; the high Laplace pressure of these small droplets makes them so
stiff that they can not be invaded by air. Indeed, when we observe small droplets
prepared by ultrasonic homogenization, we find that the small droplets are not
invaded by air but instead are simply deformed and pushed into pore space between
the surrounding particles, as shown in Figs.~4(a)-(b).

We estimate the threshold radius $R_\mathrm{th}$, beneath which droplets will not
be invaded by air by using the previous values for the parameters in
eqn.~(\ref{eqn:Delta_P}) and solving for $\Delta P = 0$; our model predicts that
$R_\mathrm{th}=2.30a=0.83 r_\mathrm{p}$. To test this prediction, we measure
$R_\mathrm{th}$ for droplets mixed with particles of several different radii,
varying $r_\mathrm{p}$ by more than half an order of magnitude. We find the
dependence $R_\mathrm{th} = (0.88\pm0.16)*r_\mathrm{p} + (0.1\pm0.1)$, in
excellent agreement with the prediction of our model, as shown in Fig.~4c.

We observe two qualitatively different behaviors: droplets that
evacuate and collapse, creating large voids; and droplets that remain
intact during drying, yielding void-free particle packs. We use thisdichotomy to produce hierarchical
materials with several different controllable length
scales, by varying droplet and particle sizes. One such structure, using
droplets with $R_\mathrm{d} \cong 20$ $\mu$m and particles with $r_\mathrm{p} =
1$ $\mu$m, is shown in Figs.~4(d)-(f). The resulting hierarchical porous
material has voids of two length scales: 20 $\mu$m, from droplets, and
0.5 $\mu$m, from the pore space between particles. Hierarchical
materials may be useful in making low-density porous materials
or to mimic hierarchical natural structures\cite{Andre,bone};
our evacuation results demonstrate drying as a general
low-energy method to drive desired materials into a porous
medium.

We gratefully acknowledge support from the NSF (DMR-0602684), Harvard
MRSEC (DMR-0213805), RGC Direct Allocation (2060395), NASA (NNX08AE09G), Pixar and ICI.

\end{document}